# Introducing the Exoplanet Escape Factor and the Fishbowl Worlds
# (Two conceptual tools for the search of extra terrestrial civilizations)

Author: Elio Quiroga Rodríguez
Universidad del Atlántico Medio, lecturer.
Las Palmas de Gran Canaria, Islas Canarias, España
elio.quiroga@pdi.atlanticomedio.es

**ABSTRACT**

The search for extraterrestrial intelligence on exoplanets is a rich field of conceptual exploration. Author introduces two definitions that could help to narrow down the possibilities that an extraterrestrial civilization may or may not have initiated exploration of its own star system and beyond. It is concluded that in some cases certain extraterrestrial civilizations may not be able to leave their home worlds to biological extinction, purely because of physical limitations.

**DISCUSSION AND PROCEEDINGS**

Within the framework of the Drake Equation, the concept of a Communicative Civilization encompasses two distinct modes of interaction: physical, involving interstellar travel (a civilization's ability to transcend the gravitational constraints of its planetary system) and informational, utilizing remote communication channels, primarily electromagnetic waves (radio). The notion of a communicative civilization may be significantly constrained by the environment in which it arises. This text explores both possibilities: space travel as a means of establishing contact with other civilizations, and also informational remote communications via electromagnetic signals.

**PHYSICAL COMMUNICATION**

About the physical communication aspect, among the articles posing possible revisions to the legendary Drake equation (Drake, F. D., 1985), some suggest variations that tilt it towards civilizations that have initiated interstellar travel, colonized other planets in their system and other star systems (Walters et al., 1980). This may lead to questions such as: How many civilizations could manage to initiate such journeys? (Benford, 2021). Is it really easy or difficult to initiate journeys within a star system, and beyond? In our human own short experience, it is not easy. The complications add up and the environment of outer space is very hostile and sterilizing (cosmic rays, solar flares, high vacuum, etc.). Could we say something about the chances of a civilization leaving its home planet?

Two papers by Michael Hippke (Hippke, 2018, 2019) on this interesting topic focus on the possible relationship between the escape velocity of an exoplanet and the possibility, or impossibility, for a possible civilization to escape from it and thus begin exploration of its own star system. In his writings, Hippke concludes that on super-Earth-like planets up to 10 times the mass of our planet, possible civilizations could escape from them by means of chemical-fuelled rockets, but for larger planetary masses, there would be no feasible possibilities or engineering to do so. Another paper (Hebbeker, 2020) goes into this detail citing Hippke's work and suggests ways for rockets to escape from planets with escape velocities greater than Earth's, pointing out the logistical (and ultimately physical) limits involving escape velocity values slightly greater than that of our planet.

It is also worth mentioning an interesting article (Gonzalez, 2020) that assesses the special "luck" in terms of probabilistic chance that the Solar System, and more specifically the Earth, has had, allowing space travel within and outside the very stellar system of which we are a part. In Figure 1 we can see the relation Gonzalez makes between the escape velocity and the mass of a super-Earth.

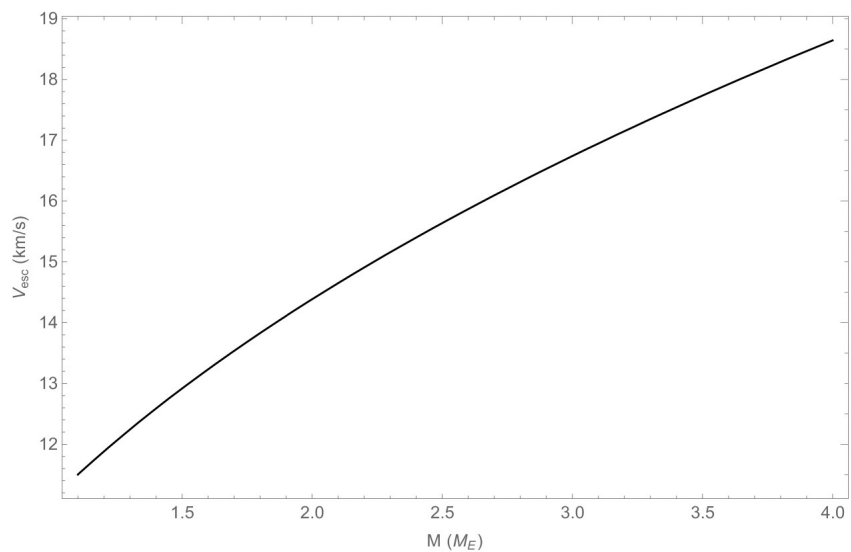

*Figure 1. Relationship of Escape Velocity and Mass as a function of Land Mass (Gonzalez, 2020).*

In his article, Gonzalez also adds to the discussion the factor of re-entry into the exoplanet of a space mission, a value that is also important, which is similar to the escape velocity, but with thermal consequences that are proportional to the square of the re-entry velocity. This indicates that any exoplanet with high escape velocities will also give problems for re-entry, with extra challenges in protecting spacecraft with heat shields (Gonzalez, 2020). All this has led the author to the following discussion which proposes to slightly systematize the issue.

The escape velocity of a planetary object is given by the equation:

$$v_e = \sqrt{\frac{2GM}{r}}.$$

The earth has a Vet (Earth's escape velocity) of 11.19 km/s, which is the minimum speed to escape the earth's gravitational grip. In the case of a planet like Jupiter with 318 times the mass of the Earth, we could be in the case of a Ve of approximately 60,2 km/s; 5,38 times greater.

Thus, the author proposes the concept of Exoplanet Escape Factor (Fex) to indicate this added difficulty for an exoplanetary civilization to leave their home world.

This would be obtained by dividing the escape velocity of the exoplanet (which we will call Vex) by the Earth's escape velocity (designated Vet, as noted above), which would lead to a value in a given range.

Fex = Vex / Vet (1)

Table 1 shows the calculations of Vex and Fex for several planets, made with equation (1) for different radii. We do not go into their composition to arrive at the masses indicated (they may be planets with a preeminence of iron, silicates, carbon, water, etc.). A color semaphore shows the difficulty to escape for a spaceship from those planets. The planets used in the table have three categories: sub-earth (Fex lower than that of the Earth), earth-like (Fex similar than that of the Earth) and super-earth (Fex bigger than that of the Earth), using values for exoplanets from Weiss & Marcy, 2014.

| Mass (T) | Radius (T) | Vex (Kms/s) | Fex | Semaphor |
|---|---|---|---|---|
| 0,85 | 2,2 | 6,95 | ,62 | Kepler-100c |
| 1,17 | 0,94 | 9,45 | ,85 | Proxima Centauri b |
| 1 | 1 | 11,19 | 1,00 | Earth |
| 2,35 | 2 | 20,90 | 1,87 | Kepler-10c |
| 6,45 | 2.65 | 16,67 | 1,49 | GJ-1214b |
| 14,11 | 3,37 | 22,89 | 2,05 | Kepler-103b |
| 16.13 | 2.41 | 26,33 | 2,35 | Kepler-131b |
| 7,34 | 1,32 | 26,38 | 2,36 | Kepler-100b |
| 10,5 | 1,71 | 27,72 | 2,48 | KOI-94b |
| 8,25 | 0,84 | 35,06 | 3,13 | Kepler-131c |

*Table 1. Comparisons of various planets with different masses, radii, Escape Velocities (Vex) and Escape Factors (Fex), with a semaphoric color code indicating the ease of exit from these celestial bodies for an eventual civilization to inhabit them (green shows a low Fex compared to Earth's, orange shows planets with possible problems for space travel, and red showing the practical impossibility of such travels) (Weiss & Marcy, 2014).*

Thus, taking into account the proposals suggested by Hippke in his articles, in the Fex range of [0.4 , 2.2] it would be reasonable that a civilization could leave its home planet (Hippke, 2019). The author does not go into the consequences cited by González in terms of heat shield requirements for the re-entry of eventual missions of exoplanetary civilizations, but it would be an interesting matter of study. In Table 1, Kepler-131c, with an Fex value of 3.13, along with KOI-94b (Fex = 2.48), Kepler-100b (Fex = 2.36), and Kepler-131b (Fex = 2.35), may not provide favorable conditions for the inhabitants to initiate space travel. Conversely, Kepler-103b (Fex = 2.05) appears to be a limit case; thus, it is likely that a prospective civilization residing on its surface would possess the capability to embark on space travel.

Values of Fex < 0.4 would cast doubt even on the possibility that the planet in question could gravitationally support an atmosphere or liquid water, at least with a radius similar to Earth's (in the case of Mars, whose seas evaporated in some past eon, its Fex is 0.45). Values of Fex > 2.2 would make space travel unlikely for the exoplanet's inhabitants: they would not be able to leave the planet using any conceivable amount of fuel, nor would a viable rocket structure withstand the pressures involved in the process (Hippke, 2019), at least with the materials we know (as far as we know, the same periodic table of elements and the same combinations of them govern the entire universe). It could therefore be the case that an intelligent species on these planets would never be able to travel into space due to sheer physical impossibility. This would have multiple consequences for their development. A civilization with a Fex > 2.2 might see space travel, even suborbital, as perhaps unconceivable. This concept of worlds that cannot or have not had the opportunity to attempt space travel due to physical impossibility has been dubbed by the author as *fishbowl worlds.*

**INFORMATIONAL COMMUNICATION**

About informational communication civilizations, let's embark on a science fiction journey for a moment, about the envision of a fishbowl world society, an extreme analogy for an oceanic or hycean world (Madhusudhan et al., 2021), where an intelligent underwater species might have thrived and evolved a civilization (however, not all fishbowl worlds would be entirely oceanic; the Fex depends on the planet's mass). Would that be a "communicative civilization" within the criteria of the Drake equation? In an underwater world imbued into a fluid, such as water or liquid methane, where sound signals can be heard hundreds of kilometers away, communication between individuals could be feasible without the need for communication devices. Also in a water world, electronics, the basis of telecommunications, would have difficulty progressing as a technology; equipment would have to be continuously isolated from water or ocean fluid. Telecommunications technology might never emerge on such a world, even though it could be home to a fully developed civilization. Such a civilization would not be "communicative" and would not be contemplated in the Drake equation.

On the other hand, while liquid water may pose challenges for the transmission of electrical signals, it does not necessarily preclude the possibility of telecommunications in an oceanic exoplanet. Scientists have been exploring the use of alternative mediums, such as plasma waves or even sound waves, for communication in underwater environments. Additionally, advances in metamaterials could enable the creation of devices that can effectively transmit signals through water.

Although telecommunications in an oceanic exoplanet; those may be challenging if the medium is water, as indicated above, because of its polarity, in the case of an oceanic exoplanet wich has liquid methane as a medium, an unpolar compound, would be possible to create circuitry in such a medium? It is also a highly corrosive and volatile substance. This could pose significant challenges for the design and fabrication of electronic devices that could operate in such an environment. Moreover, the low electrical conductivity of methane would require the development of new materi-als and technologies to ensure efficient signal transmission. Despite these challenges, it is not impossible that a civilization on an oceanic exoplanet with liquid methane as solvent could develop some form of telecommunications infrastructure.

**OTHER RELATED LIMITATIONS**

In a binary system in which it is always daylight and the stars cannot be seen, could a desire to explore an invisible planetary environment develop? And in a world perpetually covered by clouds, which prevent its inhabitants from seeing "beyond"? Or in an underwater civilization like the one the author discussed earlier, condemned to remain within the confines of its oceanic world? In addition, according to some studies, super-earths larger than 2.2 Earth radii would probably be difficult to inhabit (Alibert, 2014) for ultimately physical-biological reasons.

Worth mentioning here is an article whose main author is a disciple of Abraham Loeb's, Manasvi Lingam (Lingam et al., 2023) which contemplates a Bayesian approach seeking to model the possibility of the advent of technological intelligences on what the author has christened in this chapter as Fishbowl Worlds; planets with interior oceanic worlds (in the style of Jovian satellites such as Europa, Ganymede or Callisto[1]), which in the article they call OBHs (Ocean-Based Habitats), with a potentially high abundance in the universe as opposed to rocky planets with surface oceans, which they call LBHs (Land-Based Habitats), reaching interesting conclusions: the emergence of technological intelligences in OBH-like environments is unlikely, at least with the baseline knowledge we use: Earth, an LBH, is the only example we know of the advent of technological intelligences. That said, even though an LBH planet with a high Fex would be challenged in undertaking space travel, it could still establish telecommunications systems capable of transmitting signals to outer space, qualifying it as a potential "communicative civilization" under Drake's equation.

About electromagnetic communication, and in the case of Humanity, in a rough calculation we have been broadcasting on radio for 2.4% of our history as a civilization, using the optimistic figure of 1900 as the first use of an electromagnetic transmission for the wireless telegraph (Maver, 1912). We have been a civilization according to the scientific consensus for 97.6% of our history without making electromagnetic emissions. For this reason, perhaps our concept of "communicative civilization," as discussed in this article, at least considering the ability to be "detected" by other civilizations by our radio emissions, is conditioned by the current state of the art. On the other hand, with new developments such as the Internet, which largely operates through undersea cables, our average radio communicative emission may decrease in the future, and we may not be as "detectable" by others listening as we are now.

---

1   On Saturn, Enceladus may also contain a similar subsurface ocean.

**CONCLUSION**

The use of the so-called Exoplanet Escape Factor (Fex) is proposed as a reference to anticipate the physical capacity or inability of a planet for its inhabitants to have initiated space travel, which would lead to interesting conclusions, among which is defined that of a Fishbowl World, a planet whose physical limitations (Escape Factor, single ocean, perpetual clouds, outer crust over an ocean, etc.) would prevent the possible civilizations inhabiting it from escaping from it before their biological extinction. Contact with the inhabitants of a Fishbow World would be very difficult.

**DATA AVAILABILITY STATEMENT**

All data is available in the body of this article.